\documentclass[aps,prd,twocolumn,amsmath,amssymb,superscriptaddress,nofootinbib,showpacs]{revtex4-1}

\usepackage{graphicx}
\usepackage[breaklinks=true,colorlinks=true,linkcolor=red,citecolor=blue]{hyperref} 

\newcommand{\aj}{{AJ}}	
\newcommand{\aap}{{A\&A.}}
\newcommand{\mnras}{{MNRAS}}
\newcommand{\apjl}{{ApJ Lett.}}	

\begin{document}

\title{Constraints on MOND theory from radio tracking data of the Cassini spacecraft}

\author{A.~Hees}
\author{W.~M.~Folkner}
\author{R.~A.~Jacobson}
\author{R.~S.~Park}
\affiliation{Jet Propulsion Laboratory, California Institute of Technology, 4800 Oak Grove Drive, Pasadena CA 91109}

\date{\today}

\begin{abstract}
	The MOdified Newtonian Dynamics (MOND) is an attempt to modify the gravitation theory to solve the Dark Matter problem. This phenomenology is very successful at the galactic level. The main effect produced by MOND in the Solar System is called the External Field Effect parametrized by the parameter $Q_2$. We have used 9 years of Cassini range and Doppler measurements to constrain $Q_2$. Our estimate of this parameter based on Cassini data is given by $Q_2=(3 \pm 3)\times 10^{-27} \ \rm{s^{-2}}$ which shows no deviation from General Relativity and excludes a large part of the relativistic MOND theories. This limit can also be interpreted as a limit on a external  tidal potential acting on the Solar System coming from the internal mass of our galaxy (including Dark Matter) or from a new hypothetical body.
\end{abstract}

\maketitle
\section{Introduction}
One of the mysteries of modern astrophysics is known as the Dark Matter problem. This problem comes from the discrepancies between observations of galactic systems and the predictions of General Relativity (GR) and the Standard Model of particles. Some galactic and extragalactic dynamical observations can not be explained by GR and by the amount of observed matter.  Three possibilities can be found in the literature to solve this problem: the presence of unseen mass-energy ; a modification to the theory of gravity ; or both~\cite{famaey:2012fk}.

Modified Newtonian Dynamics (MOND) is an attempt to modify the gravitation theory to solve the Dark Matter problem. This phenomenology proposed by Milgrom 30 years ago~\cite{milgrom:1983uq,*milgrom:1983kx,*milgrom:1983fk} is very successful in explaining galactic observations (see for examples \cite{famaey:2012fk,milgrom:2008uq,sanders:2002fk} and references given in the MOND webpage~\footnote{\url{http://www.astro.umd.edu/~ssm/mond/litsub.html}}). In particular, MOND superbly explains the galactic rotation curves \cite{sanders:2002fk} and automatically recovers the Tully-Fisher law which establishes a relation between the luminosity and the rotation velocity of a spiral galaxy \cite{tully:1977kx,*mcgaugh:2000vn}.

The main idea of MOND theory is to modify the standard Newtonian gravitation law $\mathbf a=\mathbf g_N$ (where $\mathbf a$ is the acceleration of a test particle and $\mathbf g_N$ is the Newtonian gravitational field) by the relation $\mathbf a=\mathbf g$ with $\mathbf g$ determined by the relation 
\begin{equation}\label{eq:mond_g}
	\mu\left(\frac{g}{a_0}\right)\mathbf g=\mathbf g_N.
\end{equation}
Here, $\mu$ is a function of the ratio $g/a_0$ between the norm of the gravitational field $g$ and the MOND acceleration scale $a_0$. The usual value of the MOND acceleration scale $a_0$ is between $0.9 \times 10^{-10}\ \rm{m/s^2}$  and $1.2\times 10^{-10} \ \rm{m/s^2}$ \cite{gentile:2011uq}.  In order to recover the Newtonian dynamics in a strong gravitational field (formally when $g>>a_0$), the function $\mu$ needs to satisfy $\mu(x)\to 1$ when $x\to\infty$. The MOND regime appears in the limit of weak gravitational field $g<<a_0$ where the interpolating function has to satisfy $\mu(x)\to x$ when $x\to 0$ in order to explain galactic rotation curves~\cite{milgrom:1983uq,milgrom:1983kx,milgrom:1983fk}. Various $\mu$ functions are possible interpolating between the MOND and the Newtonian regimes. 

Relativistic extensions of MOND have been developed and are more satisfactory from a theoretical point of view. The Bekenstein Tensor-Vector-Scalar (TeVeS) theory was the first relativistic extension of MOND~\cite{sanders:1997kx,*bekenstein:2004fk,*sanders:2005fk}.  These relativistic extensions have now evolved into Einstein-Aether theories~\cite{jacobson:2001qf,*zlosnik:2006dq,*zlosnik:2007bh,*zhao:2007vn}. Very detailed reviews of relativistic MOND theories can be found in~\cite{bruneton:2007vn,famaey:2012fk}. Finally, a new interpretation of MOND in term of dipolar dark matter has been developed in~\cite{blanchet:2007kx,*blanchet:2008fv,*blanchet:2008vn,*blanchet:2009kx}.

While being very successful at the galactic level, the modification of the gravitation theory provoked by MOND has to be small enough in the Solar System to be able to recover the well known Solar System dynamics. Within the Solar System, three types of effects can arise due to MOND, the most important one being the MOND External Field Effect\footnote{We use the term External Field Effect as in \cite{blanchet:2011ys} to denote this effect that should not be confused with the fact that large external gravitational field may influence the internal dynamics directly because it is the total acceleration that enters the $\mu$ function \cite{milgrom:1983fk}} \cite{milgrom:2009vn,blanchet:2011ys,blanchet:2011zr}.

The EFE produces a quadrupole correction (parametrized by $Q_2$) to the Newtonian potential which increases with the distance to the Sun.
As shown in \cite{blanchet:2011ys,blanchet:2011zr}, Saturn is the Solar System body that is most likely to allow for detection of the EFE, since the orbit is relatively far from the Sun but with orbit period short enough to allow separation of initial conditions from dynamical model parameters. 
In order to constrain MOND EFE, we consider below radio tracking measurements of the Cassini spacecraft, taking care to account for systematic correlations between measurements and the orbits of the spacecraft and the orbit of Saturn.
We have added the MOND EFE potential to the dynamical model of the orbits of the planets and have estimated the $Q_2$ parameter in the global fit. With this approach, we derive an estimate of the parameter characterizing the MOND EFE that constrains very severely the MOND theory. Moreover, the quadrupolar potential parametrized by $Q_2$ can also be produced by a tidal interaction. Therefore, our constraint on $Q_2$ can also be interpreted as a constraint on the internal mass (including Dark Matter) of our galaxy \cite{braginsky:1992uq,klioner:1993fk}  or as a constraint on the mass of a hypothetical new body \cite{iorio:2012rm,*iorio:2010yg}.

\section{MOND in the Solar System}\label{sec:mondss}
The relation (\ref{eq:mond_g}) can be explained by three different types of phenomenology. 
\begin{itemize}
	\item The first one is called modified inertia. In this approach, the gravitational field potential is still determined by the Newtonian Poisson equation but the particle equations of motion are modified~\cite{milgrom:1994ve,milgrom:2011fk}. A serious drawback of this approach is that these theories are non-local~\cite{felten:1984uq}. 
	\item The second and most widespread approach consists in modifying the Newtonian Poisson equations. In this approach proposed by \cite{bekenstein:1984kx}, the modified Poisson equation takes the form
	\begin{equation}\label{eq:bekenstein}
		\nabla \cdot \left[\mu\left(\frac{\left|\nabla \Phi\right|}{a_0}\right)\nabla\Phi\right]=4\pi G\rho=\nabla^2\Phi_N
	\end{equation}
	with $G$ the Newton constant, $\rho$ the matter density, $\Phi_N$ the Newtonian gravitational potential solution of the classical Poisson equation and $\Phi$ the gravitational potential from which the equations of motion are derived $\mathbf a=-\nabla \Phi$.
	\item The third recently developed approach is called quasi-linear MOND (or QUMOND)~\cite{milgrom:2010uq}. In QUMOND, the gravitational field is the solution of the equation
	\begin{equation}\label{eq:qumond}
		\nabla^2\Phi=\nabla\cdot\left[\nu\left(\frac{|\nabla\Phi_N|}{a_0}\right)\nabla\Phi_N\right].
	\end{equation}
	This formalism requires solving the Newtonian linear Poisson equation to determine $\Phi_N$ and then solving the above Poisson equation to find $\Phi$ with $\nu$ being another kind of interpolating function.
\end{itemize}  
It should be noted that the Bekenstein-Milgrom and QUMOND approaches coincide in the case of spherical situations but are not equivalent in non symmetric cases~\cite{milgrom:2010uq,zhao:2010uq}. In the case of spherical symmetry, the interpolating  functions  $\mu$ and $\nu$ are related by $\nu(y)=1/\mu(x)$ with $x$ and $y$ related by $x\mu(x)=y$~\cite{milgrom:2010uq}.

These phenomenologies mainly produce three different kinds of effects in the Solar System.
\begin{itemize}
	\item The first effect comes from a departure of the MOND interpolating function ($\mu$ or $\nu$) from unity  and produces a deviation from the Newtonian gravity able to explain the rotation curves of disk galaxies. As can be seen from (\ref{eq:bekenstein}) and (\ref{eq:qumond}), the anomalous acceleration produce by MOND in the case of a spherical system can be written as
	\begin{equation}
		\mathbf \delta \mathbf a=\left[\nu\left(\frac{g_N}{a_0}\right)-1\right]\mathbf g_N.
	\end{equation}
 This deviation highly depends on the MOND interpolating function used. Various MOND interpolating functions exist in the literature (see \cite{famaey:2012fk} for a review). The most widely considered functions are:
\begin{subequations}
	\begin{eqnarray}	
		\mu_n(x)&=&\frac{x}{\sqrt[n]{1+x^n}}\label{eq:mu1}\\
		\mu_{\rm exp}(x)&=&1-e^{-x}\\
		\mu_{\rm TeVeS}(x)&=&\frac{\sqrt{1+4x}-1}{\sqrt{1+4x}+1}\label{eq:muteves}\\
		\mu(x)&=&\frac{\sqrt{1+4x^2}-1}{2x}.
	\end{eqnarray}
\end{subequations}
Solar System constraints on this effect have been studied in~\cite{milgrom:1983fk,sereno:2006vn,iorio:2008qv}. In particular, this effect can be made arbitrarily small by suitable choice of the interpolating function (for example, $\mu_{\rm exp}$ produces an undetectable deviation from the Newton equations of motion). 

\item The second effect called the Solar System External Field Effect is produced by MOND theories of gravitation based on nonlinear extensions of the Poisson equation. This effect appears even for arbitrarily fast vanishing $\mu-1$ functions corresponding to a fast transition to the Newtonian regime. It is due to the non-linearity of MOND equations in which the gravitational dynamics of a system are influenced by the external gravitational field. This effect has been studied in the framework of the Bekenstein modification of the Poisson equation~(\ref{eq:bekenstein}) in~\cite{milgrom:2009vn,blanchet:2011ys}. It implies the presence of an anomalous quadrupolar correction to the Newtonian potential

\begin{eqnarray}\label{eq:efe}
	\Phi&=&-\frac{GM}{r}-\frac{Q_2}{2}x^ix^j\left(e_ie_j-\frac{1}{3}\delta_{ij}\right)\\ &=&\Phi_N+\delta\Phi_N \nonumber
\end{eqnarray}

where $e_i$ is a unitary vector pointing towards the galactic center and $\Phi_N$ is the classical Newtonian potential. 

The EFE arising in QUMOND is also of the form of a quadrupolar correction to the Newton potential \cite{milgrom:2010uq}. The main difference between the EFE in Bekenstein-Milgrom theory and the EFE in QUMOND is that the unit vector $e_i$ is pointing in the direction of the Newtonian galactic field in QUMOND while it is pointing towards the MOND galactic gravitational field in the non-linear Poisson theory. This effect is not present in the Modified Inertia approach to MOND theory \cite{milgrom:2011fk}.

The value of the quadrupole $Q_2$ can be computed from the theoretical model of MOND and depends on the MOND interpolating function and on the ratio $\eta$ between the external gravitational $g_e$ field and the MOND acceleration $a_0$. Let us mention that in \cite{blanchet:2011ys}, the value of $Q_2$ has been determined numerically for the MOND interpolating functions (\ref{eq:mu1}-\ref{eq:muteves}) with $g_e=1.9\times 10^{-10} \ m/s^2$ and $a_0=1.2\times 10^{-10} \ m/s^2$. The obtained values for $Q_2$ are bounded by two limits

\begin{equation}\label{eq:Q2}
	2.1 \times 10^{-27} \ \rm{s^{-2}} \leq Q_2 \leq 4.1 \times 10^{-26} \ \rm{s^{-2}} 
\end{equation}
depending on the MOND function used. Note that in \cite{milgrom:2009vn}, the potential (\ref{eq:efe}) is parametrized by 
\begin{equation}\label{eq:q}
	q=-2Q_2r_M/3a_0
\end{equation} 
(with $r_M=(GM/a_0)^{1/2}$ the MOND radius). The value of this parameter is computed theoretically for different MOND interpolating function in \cite{milgrom:2009vn}.  These values are sensitive to $\eta$. A small change of the external galactic gravitational field $g_e$ or of the MOND acceleration $a_0$ may change substantially the predicted values of $q$. The estimated value of $g_e$ is usually between $1.9\times 10^{-10} \ m/s^2$ and $2.4\times 10^{-10} \ m/s^2$ \cite{mcmillan:2010jk} while the value of the MOND acceleration is between $a_0=0.9 \times 10^{-10} \ m/s^2$ and $a_0=1.2 \times 10^{-10} \ m/s^2$ \cite{gentile:2011uq}. This means the value of $\eta$ is therefore between $\eta=1.6$ and $\eta=2.7$.

Finally, it is worth mentioning that the quadrupole potential (\ref{eq:efe}) is not a feature of MOND only. In particular, this modification of the Newtonian potential is also produced by the tidal perturbation coming from a third body \cite{kaula:1966kx}. Therefore, a similar potential is produced by the Newtonian tidal interaction coming from a "new" planet. This effect has been investigated in \cite{iorio:2012rm,*iorio:2010yg}. On the other hand, a similar potential is also produced by the Newtonian tidal interaction coming from our galaxy \cite{braginsky:1992uq,klioner:1993fk}. In the spherically symmetric approximation, the $Q_2$ parameter is given by 
\begin{equation}\label{eq:tidal}
	Q_2=3GM(D)/D^3
\end{equation} 
with $D$ the distance between the Solar System and the galactic centre and $M(D)$ the total mass enclosed in a sphere of radius $D$. The contribution of the stellar mass of our galaxy gives $Q_2\sim 10^{-30} \ \rm{s^{-2}}$ \cite{klioner:1993fk} while a Dark Matter density of $\rho_{\textrm{DM}}=0.4 \; \rm{GeV/cm^3}$ \cite{mcmillan:2011vn} gives $Q_2=6 \times 10^{-31} \ \rm{s^{-2}}$. The Newtonian tidal interaction from our galaxy is therefore smaller than the effects predicted by the EFE of MOND.

\item The last effect is also produced by modified Poisson equation in the Solar System and results from the fact that the Solar System is aspherical \cite{milgrom:2012fk}. In an aspherical system, a quadrupolar contribution to the Newton potential appears in the framework of QUMOND
\begin{equation}
		\Phi=-\frac{GM}{r}-\frac{G\alpha}{r_M^5}Q_{ij}x^ix^j
\end{equation}
with $Q_{ij}$ the quadrupolar moment of the Solar System computed in the Solar System barycentric frame, $\alpha$ a constant depending on the MOND interpolating function and $r_M=\left(GM/a_0 \right)^{1/2}$ the MOND radius ($\sim 10^{12}\ \rm{ km}$ for the Sun). A similar effect is expected in the Bekenstein-Milgrom approach but its exact expression has not been computed yet \cite{milgrom:2012fk}. As mentioned in \cite{milgrom:2012fk,iorio:2013ty}, the order of magnitude of this effect is below present measurement capabilities.
\end{itemize}

\begin{figure}[htb]
\begin{center}
\includegraphics[width=0.5\textwidth]{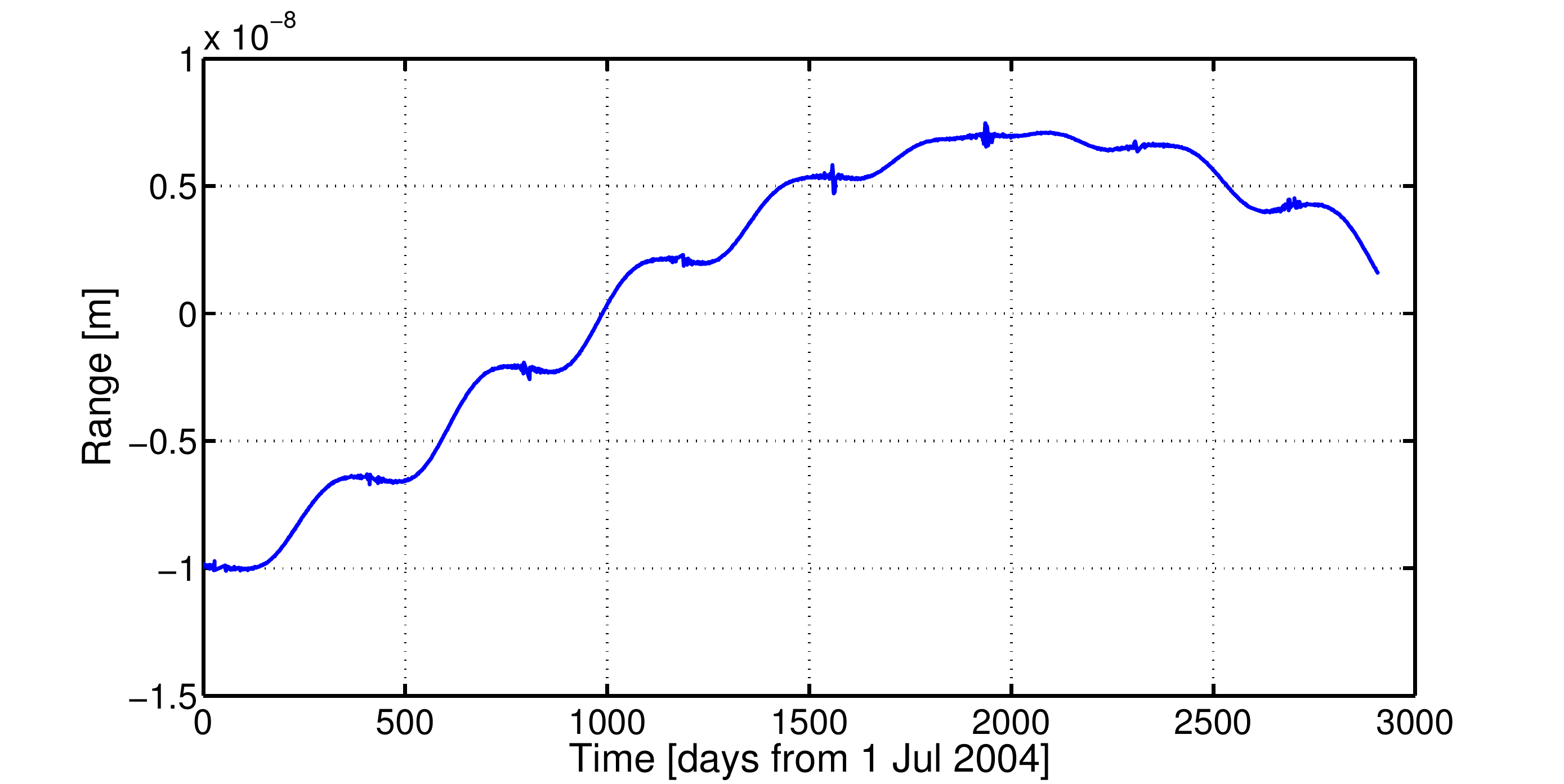}
\end{center}
\caption{Representation of the effect of the modification of the light propagation due to the $\delta\Phi_N$ term in metric~(\ref{eq:metric}) on the Earth-Saturn range.} 
\label{fig:time}
\end{figure}

Since the first effect can be made arbitrarily small by the choice of the interpolating function and since the last effect is below our present detecting capabilities, we concentrate on the External Field Effect (EFE). Therefore, we consider a modification of the Newton potential of the form~(\ref{eq:efe}) that has an obvious consequence on the equations of motion for bodies. In addition, the effects  on the light propagation coming from a relativistic extension of MOND should be considered. Therefore, it is better to consider a modification of the space-time metric related to complete and well defined relativistic MOND theories. The metric

\begin{eqnarray}\label{eq:metric}
	ds^2&=&-\left(1-2\Phi_N-2\delta\Phi_N + 2 \Phi_N^2 \right)dt^2 \\&&\quad+ (1+2\Phi_N+2\delta\Phi_N)\delta_{ij}dx^idx^j \nonumber
\end{eqnarray}
is the low field metric derived from the TeVeS theory \cite{bekenstein:2004fk,skordis:2008bh} or from certain Einstein-Aether theories \cite{zlosnik:2006dq,zlosnik:2007bh}. The metric~(\ref{eq:metric}) allows one to compute not only the equations of motion of bodies (planets or spacecraft) but also the light propagation. The effect of the alternative theory on the light propagation has to be considered in the analysis of spacecraft data. In the case of the relativistic MOND extension characterized by the metric~(\ref{eq:metric}), the modification of the light propagation on the Earth-Saturn range (the modification of the Shapiro delay) is completely negligible. Fig.~\ref{fig:time} represents the effect of the modification of the light propagation due to the additional quadrupolar term in the metric~(\ref{eq:metric}) on the Earth-Saturn range. This figure was obtained using software presented in \cite{hees:2012fk} based on the time transfer formalism \cite{hees:2014fk} and shows the effect of the modification of the light propagation on the range is always smaller than $10^{-8}$ \rm{m} (or equivalently $3\times 10^{-16}$ \rm{s}), far below present range measurement accuracy.

The sensitivity of the MOND EFE in the Solar System has been studied in \cite{milgrom:2009vn,blanchet:2011ys,blanchet:2011zr}. In particular, the anomalous perihelion precession caused by the MOND EFE has been compared to published postfit residuals for any possible supplementary precession of planetary orbits derived from the INPOP ephemerides~\cite{blanchet:2011ys,fienga:2011qf}. Nevertheless, as mentioned in~\cite{blanchet:2011zr}, the INPOP ephemerides are used to detect the presence of an eventual abnormal precession, not to adjust precisely the value of that precession and the postfit residuals are obtained by adding {\it ad hoc} excesses of precession for the planets and constraining these excess by looking at the way the postfit data change~\cite{fienga:2011qf}. In this study, we work completely in a MOND framework by integrating the MOND equations of motion. This results in other effects produced by the theory like variations of the eccentricity and of the inclination~\cite{blanchet:2011ys,blanchet:2011zr}. Then, we perform a global fit of all these effects to the data. 

\section{Cassini data reduction}
The Cassini spacecraft trajectory involves a series of highly elliptical orbits about Saturn designed to give multiple close approaches to Saturnian satellites and varying views of Saturn's rings \cite{wolf:1995ys,strange:2002ve}. The changes in trajectory from one orbit to the next are made utilizing gravity assists from the satellite encounters and a series of propulsive maneuvers. Radio Doppler and range measurements between the spacecraft and tracking stations of the NASA Deep Space Network (DSN) are used to determine the spacecraft orbit and to estimate the gravity fields of Saturn and its satellites \cite{antreasian:2005kx,jacobson:2006kx,iess:2010ys,iess:2012vn}. The radio measurements during satellite encounters are also used along with imaging of the satellites against a background star field to estimate the orbits of the satellites about Saturn.

In order to use the range measurements to the Cassini spacecraft to estimate the orbital motion of the Saturnian system about the Sun, the position of the spacecraft with respect to the Saturn system barycenter must be determined. The spacecraft trajectory is produced by numerical integration of the equations of motion. The equations of motion are formulated in Cartesian coordinates referred to the International Celestial Reference System (ICRS), realized by the extragalactic radio positions which define the International Celestial Reference Frame 2 (ICRF2) \cite{fey:2009fk}. 

The forces acting on the spacecraft include the point-mass Newtonian accelerations due to the Sun, the planets, and the Saturnian satellites; the relativistic perturbations due to the Sun, Jupiter, and Saturn; and the perturbation due to the oblateness of Saturn. Details of the equations may be found in \cite{moyer:2000uq}. The spacecraft is also subject to a variety of nongravitational forces from trajectory correction maneuvers, attitude control maneuvers, solar radiation pressure, and non-isotropic thermal radiation from the radioactive isotope thermal generators that provide electrical power for the spacecraft.

No modification due to the External Field Effect has been taken into account in the spacecraft orbit determination. This is justified for three reasons. First of all, the spacecraft orbit is determined with respect to the Saturnian barycenter while the additional EFE force is centered on the Sun. Therefore, the effect of the additional force on the position of the spacecraft relative to the Saturnian barycenter is a differential effect that is much smaller with respect to the EFE on Saturn and therefore is completely negligible. Moreover, the length of the typical orbit segments of the spacecraft orbit reconstruction is a few days due to the numerous maneuvers and encounters. In this time, the effect of the modification of the gravitation theory does not have the time to accumulate. Finally, as mentioned above (see Fig.~\ref{fig:time}), the modification of light propagation due to the EFE is completely negligible. Therefore, we can safely neglect the effect of MOND for the spacecraft orbit determination.

\begin{figure}[htb]
\begin{center}
\includegraphics[width=0.9\linewidth]{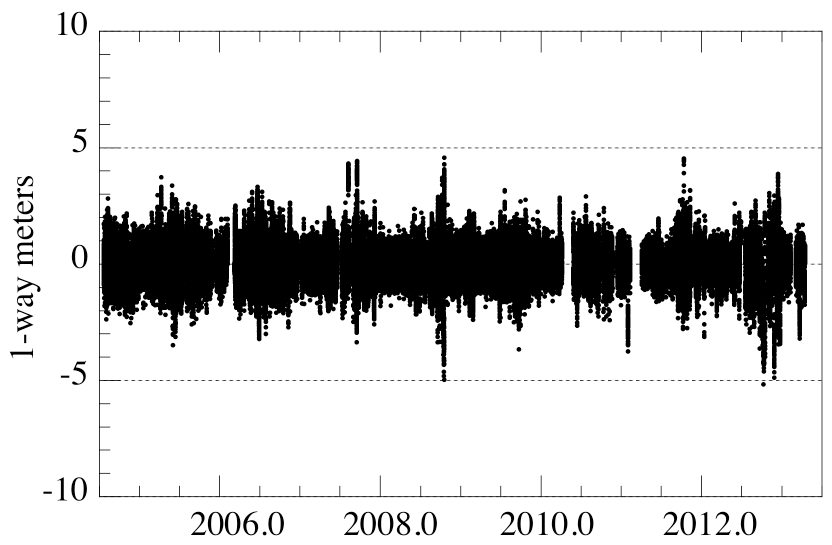}
\end{center}
\caption{Representation of range measurement residuals (one per orbit segment).} 
\label{fig:cass_Ran_fit}
\end{figure}

Estimation of the spacecraft orbits involves estimation of a number of parameters describing the nongravitational forces. With current models, the range and Doppler can be fit to their intrinsic noise level (0.75 meters for the range and 0.1 \rm{mm/s} for the Doppler) without any signature remaining when using both range and Doppler (see Fig.~\ref{fig:cass_Ran_fit} for the range residuals ; a similar behavior is found for Doppler). This is due to the number of free parameters that are estimated (in particular the numerous maneuvers). For this study, we have estimated spacecraft trajectories with only Doppler and satellite imaging data. With this method, the Doppler is also fit to its intrinsic noise level. Omitting the range data from the spacecraft trajectory estimates leads to larger range residuals since they are not absorbed in the spacecraft orbit parameters but this allows the range data to be used to estimate corrections to the Saturnian orbit. 

For our orbits estimated without range measurements, the maneuvers are not well determined. As a result the spacecraft orbits are effectively broken into shorter segments, with duration defined by either a maneuver, a satellite encounter, or a Saturn pericenter passage. The typical length of an orbit segment is between 15 and 30 days before mid 2009 with about one DSN pass per day. After mid 2009, the orbit period is shorter and the typical length of an orbit segment is 10 days. 

For each orbit segment the initial spacecraft position and velocity were estimated along with corrections to spacecraft maneuvers and nongravitational accelerations, the orbits of the Saturnian satellites, the right ascension an declination of Saturn's pole, the Saturn mass parameter $GM$, Saturn zonal harmonics $J_2$, $J_4$, and $J_6$ as described in \cite{jacobson:2006kx}. The nominal values for these dynamical parameters were taken from a fit of all ten years of tracking data since entering orbit about Saturn. Estimated uncertainties for the satellite orbit parameters and Saturn gravity field from the long-arc fit were used as a priori uncertainties for fitting the spacecraft initial position and velocity for each segment. Orbit estimation with primarily Doppler data does not determine all orbital elements equally accurately. In particular, the orientation of the orbit about the direction between Earth and Saturn (line of nodes) is relatively poorly estimated~\cite{wood:1986fk}. For our purposes the primary orbital uncertainty of interest is the distance from the spacecraft to the Saturn system barycenter, and this is insensitive to rotation of the orbit about the direction from Earth to Saturn.

The radio Doppler measurements as implemented in the DSN actually measure the change in distance (round-trip light time divided by the speed of light) between the tracking station and the spacecraft during the measurement interval by measuring changes in the radio carrier phase delay \cite{thornton:2000uq}. An X-band carrier at $7.2\; \rm{GHz}$ is sent from a DSN station to the spacecraft. The onboard transmitter retransmits the signal back to Earth at X-band (8.4 \rm{GHz}). In addition, a Ka-band (32.5 \rm{GHz}) downlink is used for some satellite encounters. Typically for Doppler measurement interval (count time) of 60 seconds, the change of distance divided by the measurement time is accurate to $<0.1 \; \rm{mm/s}$ \cite{asmar:2005fk}. Doppler measurement accuracy, at the radio frequencies used for Cassini, is largely limited by random fluctuations in the integrated number of electrons between the tracking station and the spacecraft. 

\begin{figure}[htb]
\begin{center}
\includegraphics[width=0.8\linewidth]{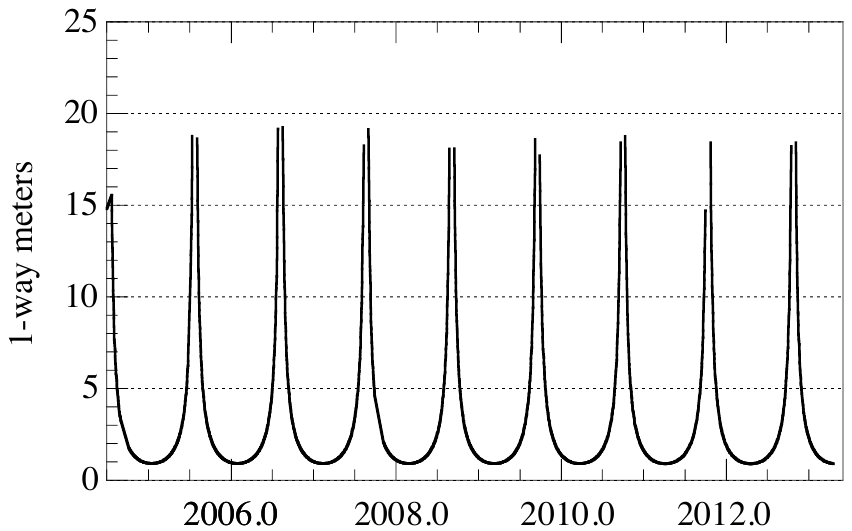}
\end{center}
\caption{Plasma contribution to the Earth-Cassini range measurements.} 
\label{fig:plasma}
\end{figure}

\begin{figure}[htb]
\begin{center}
\includegraphics[width=0.8\linewidth]{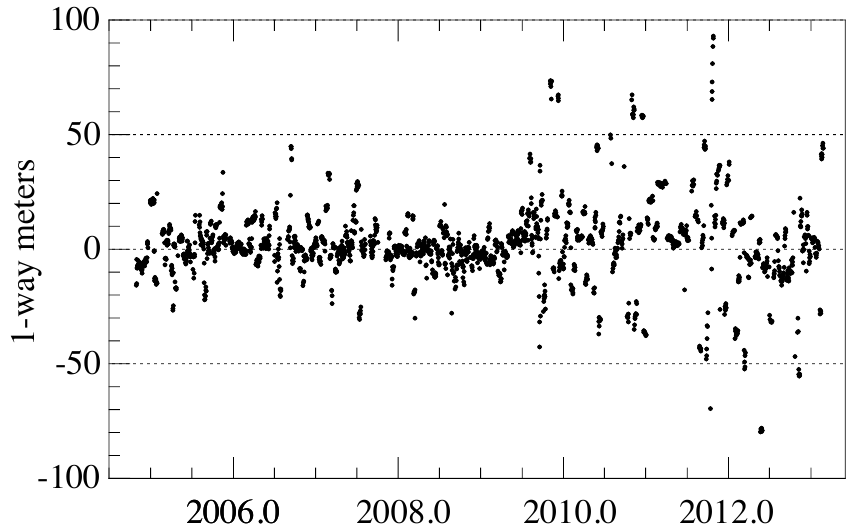}
\includegraphics[width=0.8\linewidth]{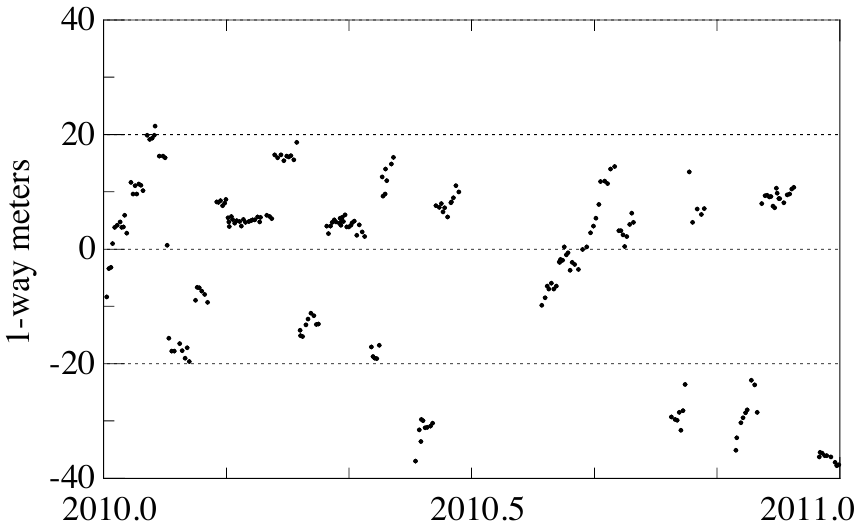}
\end{center}
\caption{Cassini range measurement residuals (one per tracking pass) when using Doppler data only to fit the spacecraft orbit (the second plot is a zoom of the first one).} 
\label{fig:res1}
\end{figure}

Range measurements determine the distance to the spacecraft using a modulation on the radio carrier signal. The range measurement accuracy of 75 centimeters is limited by calibration of the signal delay of the DSN tracking station and electronics measured at the start of each tracking pass \cite{kuchynka:2012kx}. Because this calibration error is common to each range measurement during the tracking pass, and the changes in range are measured by the Doppler measurements with great accuracy, there is essentially only one independent range measurement for each tracking pass. Range measurements are processed by using the method described in \cite{moyer:2000uq}. The round-trip light time is calculated based on the positions of Earth and Saturn barycenter from the nominal ephemeris integration, the spacecraft trajectories estimated without use of range data, and standard models for Earth rotation. Calibrations to the measured light time are applied for the tracking station path delay, the spacecraft radio delay, and the effects of the Earth troposphere and ionosphere. A nominal model for signal delay due to solar plasma has also been applied \citep{standish:1990vn,asmar:2005fk} with a constant scale factor correction estimated that allows for a possible variation of the average particle density. The amplitude of the solar plasma delay is shown on Fig.~\ref{fig:plasma}.  A plot of the range measurement residuals is shown in Fig.~\ref{fig:res1}. The range measurements include points as close as 1.9 degrees in angle from the Sun (we use the angle Sun-Earth-planet).

\section{Analysis and results}\label{sec:results}
For estimating the corrections to the orbit of Saturn and the EFE, the orbits of the Sun and planets have been integrated using the equations of motion given by \cite{standish:2012fk} plus the EFE effect from Equation (\ref{eq:efe}).

The Cassini range data are sensitive to the orbit of the Earth, the mass parameter ($GM$) of the Sun, the initial position and velocity of Saturn, and the $Q_2$ parameter. The orbit of the Earth and the mass parameter of the Sun are well determined from radio ranging and very-long-baseline interferometry (VLBI) measurements of spacecraft in orbit about Mars  \cite{folkner:2009fk,konopliv:2011dq,folkner:2013uq}. For estimation of the MOND EFE we have held the orbit of the Earth and mass parameter of the Sun fixed to the values from the fit from the planetary ephemeris DE430 \cite{folkner:2014uq}. We have used the Cassini range measurements and VLBI observations of Cassini to estimate corrections to the 6 Saturnian orbital elements and the $Q_2$ parameter scaling the MOND EFE effect.

In addition to these dynamical parameters, the radio range is affected by an instrumental delay in the spacecraft radio. The spacecraft radio range delay is calibrated prior to launch but may change slightly due to radiation, temperature changes, and aging of electronic components in the spacecraft radio system. A constant correction to the radio delay has been estimated to account for these effects. The nominal value of the radio delay is 0.42~$\rm{\mu s}$ with a constraint of $\pm$~10~\rm{ns}. 

The 12 VLBI measurements of the Cassini spacecraft included in the fit constrain the orientation of the orbit of Saturn and are insensitive to the MOND EFE. The accuracy of the VLBI data is 0.5 milliarcsecond which correspond to an accuracy of 7 \rm{km} at Saturn distance. VLBI residuals can be found in \cite{jones:2011fk}.

\begin{figure}[htb]
\begin{center}
\includegraphics[width=0.7\linewidth]{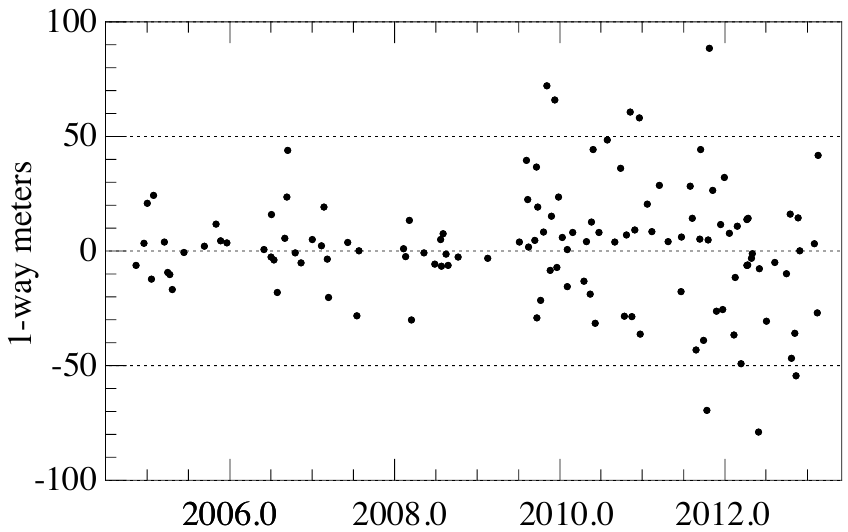}
\end{center}
\caption{Representation of range measurement residuals (one per orbit segment).} 
\label{fig:res}
\end{figure}

Concerning the weighting of the observations in the least-square parameters estimation, the Gauss-Markov theorem guarantees that the method is optimal when observation errors are independent and the observations are weighted by the square roots of their individuals covariances. As mentioned above, the range measurements within a single tracking pass are correlated in the sense that they share a common error at the DSN station. Therefore, treating each range measurement as statistically independent from the others significantly over-estimates the accuracy with which parameters can be estimated (see also e.g. Sect.~11 of \cite{konopliv:2011dq}). An initial analysis was performed treating one range measurement from each tracking pass as independent. An assessment of estimates made with different subsets of the data indicated that the approach resulted in estimate variations much larger than the associated uncertainties. This is due to the fact that ranging measurements from a single spacecraft orbit segment are correlated because they share a common spacecraft orbit error. This can be seen in Fig.~\ref{fig:res1} where the range residuals appear in groups, with the change in residuals from one group to another corresponding to a change in the spacecraft orbit segment.  In order to obtain realistic uncertainties on the estimated parameters, we considered only one ranging measurements per spacecraft segment as independent. The range residuals (one per orbit segment) used for the determination of the $Q_2$ parameters are presented in Fig.~\ref{fig:res} and the spacecraft data corresponding to this plot are available as supplement material. The residuals obtained using one point per orbit segment is shown on Fig.~\ref{fig:res}.

To test the validity of the estimated parameters and uncertainties we performed cross-validation tests by considering different subsets of the data. We divided the data set into three independent subsets. Subset 1 covers the period between May 2004 and October 2006 ;  Subset 2 covers the period between October 2006 and May 2009 ; and Subset 3 covers the extended phase of the mission from May 2009 up to April 2013. We estimated the parameters described above with various combinations of these subsets. From the 166 orbit segments, we discarded data from 35 segments which had range residuals larger than 180 meters, which is more than 6 times the root-mean-square residual of the remaining 131 orbit segments. The outlying points are due to unusually large spacecraft orbit errors, which can occur for orbit segments with little or no tracking data or poor tracking geometry. The remaining data were weighted at the root-mean-square level of the three subsets; 15 meters for the 24 points in subset 1, 11 meters for the 23 points in subset 2, and 30 meters for the 84 points in Subset 3. Subset 3 contains more points with larger variance than the first two subsets because during that time the Cassini spacecraft orbit period about Saturn resulted in shorter, more numerous orbit segments with fewer Doppler tracking measurements per segment.

\begin{figure}[hbt]
\begin{center}
\resizebox{\hsize}{!}{\includegraphics{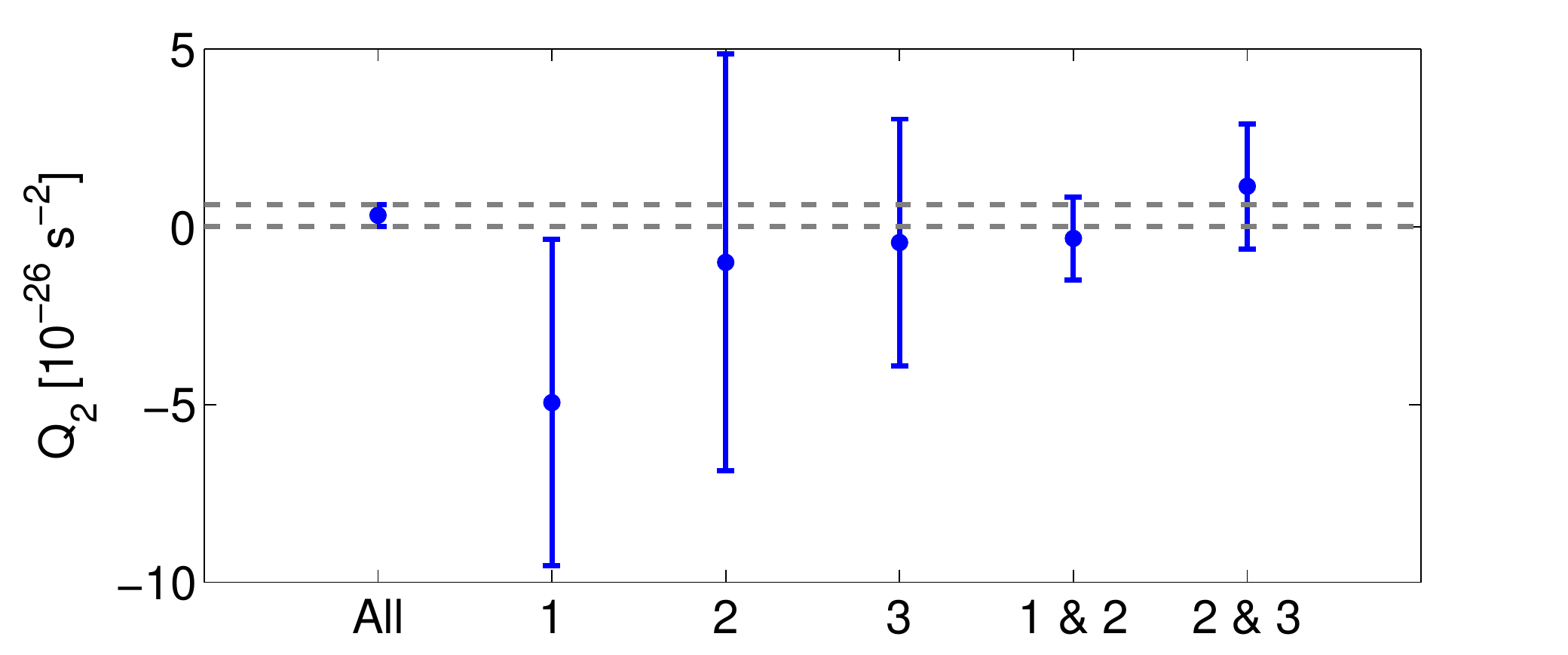}}
\end{center}
\caption{Representation of the estimation of the $Q_2$ parameter and its related 1-$\sigma$ uncertainty as function of the subset of data used.} 
\label{fig:Q2}
\end{figure}

The results of these estimates and uncertainties of the $Q_2$ parameter from different combinations of the three data subsets are presented on Table~\ref{tab:est}. The same information is shown graphically in Fig.~\ref{fig:Q2}. The estimated corrections to the orbit of Saturn, the transponder delay, and the solar corona scaling factor are less than their estimated uncertainties.

\begin{table}[htb]
\caption{Estimations of $Q_2$ and related uncertainties based on different subsets of the Cassini data.}
\label{tab:est} 
\centering
\begin{tabular}{c c c}
\hline
Set of data & $Q_2 [10^{-26} \ \rm{s^{-2}}]$  & $\sigma_{Q_2} [10^{-26} \ \rm{s^{-2}}]$\\
\hline
All data & \phantom{-}0.3 & 0.3 \\
Subset 1 & -4.9 & 4.6 \\
Subset 2 & -1.0 & 5.9 \\
Subset 3 & -0.4 & 3.5 \\ 
Subset 1 \& 2& -0.3  & 1.2 \\
Subset 2 \& 3& \phantom{-}1.1  & 1.8 \\
\hline
\end{tabular}
\end{table}

From Table~\ref{tab:est} or Fig.~\ref{fig:Q2}, we can see that all the 2-$\sigma$ confidence intervals overlap and that nearly all the 1-$\sigma$ confidence regions overlap. This is a good indication that the results and uncertainties are robust. 

We obtain an estimate of the $Q_2$ parameter given by
\begin{equation}\label{eq:Q2est}
	Q_2=(3 \pm 3)\times 10^{-27} \ \rm{s^{-2}}.
\end{equation}

\section{Discussion of the result and precession of Saturn perihelion}\label{sec:per}
The estimation (\ref{eq:Q2est}) can be transformed into an estimation on the $q$ parameter (\ref{eq:q}) used in \cite{milgrom:2009vn}. Using a value of $a_0=1.2\times 10^{-10} \ \rm{m/s^2}$ (corresponding to $\eta=\frac{g_e}{a_0}\approx 1.6$), the previous constraint translates
\begin{equation}\label{eq:qest1}
	-q=0.017 \pm 0.017
\end{equation}
while using a value of $a_0=0.9\times 10^{-10} \ \rm{m/s^2}$ \cite{gentile:2011uq} (corresponding to $\eta\approx 2.5$)  leads to a constraint
\begin{equation}\label{eq:qest2}
	-q=0.027 \pm 0.027.
\end{equation}
The value $Q_2=q=0$ is included in the 1-$\sigma$ confidence interval. This means the set of data used does not favor a MOND theory with respect to GR. Moreover, our result (\ref{eq:Q2est}) puts a very stringent constraint on the interval~(\ref{eq:Q2}) computed theoretically. The results (\ref{eq:qest1}-\ref{eq:qest2}) can be directly compared to the computed values given in Table~1 of \cite{milgrom:2009vn}. In particular, a MOND characterized by standard MOND interpolating functions like $\mu_{1,2,3}$, $\mu_{\rm exp}$ or $\mu_{\rm TeVeS}$ (see~\cite{famaey:2012fk} for a review of the MOND interpolating functions) are excluded by Cassini data. On the other hand, interpolating functions like $\mu_\infty$ or $\bar \mu_2$ are still acceptable.

As mentioned in Sec.~\ref{sec:mondss}, the potential (\ref{eq:efe}) is also produced by the tidal perturbation coming from a third body. In particular, the tidal interaction of our galaxy (the stellar mass as well as the Dark Matter) will have a similar expression. This means our constraint on $Q_2$ can be interpreted as a constraint on the internal mass of our galaxy. Nevertheless, the theoretical value of $Q_2$ coming from the Newtonian perturbation of the galaxy is three orders of magnitude smaller than our constraint (\ref{eq:Q2est}) as can be seen from the values given after Eq. (\ref{eq:tidal}) \cite{klioner:1993fk,mcmillan:2011vn}. This shows that the effects produced by the tidal interaction coming from our galaxy is  too small to be currently detected in Solar System observations.

Furthermore, our estimation (\ref{eq:Q2est}) can be used to constrain the mass and the distance of a new body in the Solar System through the third body tidal interaction. For example, in \cite{trujillo:2014uq}, the existence of a new super massive Earth (2-15 Earth masses located between 200 and 300 $AU$) has been suggested. The constraint (\ref{eq:Q2est}) implies that if the new massive body is currently located in the direction of the galactic center, its distance from the Sun has to be larger than 490 $AU$ for a 2 Earth mass body and larger than 960 $AU$ for a 15 Earth mass body. A more complete analysis is done in \cite{iorio:2014fk}.

As computed in \cite{blanchet:2011ys,blanchet:2011zr}, the quadrupolar correction to the Newtonian precession (\ref{eq:efe}) induces a precession of the perihelion denoted by
\begin{equation}
	\Delta_2=\frac{Q_2\sqrt{1-e^2}}{4n} \left[1+5\cos (2\tilde \omega)\right]
\end{equation}
where $e$ is the orbital eccentricity, $n$ is the mean motion and $\tilde \omega$ is the azimuthal angle between the direction of the perihelion and that of the galactic centre. The constraints on the $Q_2$ parameter (\ref{eq:Q2est}) can therefore be transposed in terms of a constraint on the  precession of Saturn perihelion
\begin{equation}
	\Delta_2= (0.43\pm0.43) \ \rm{mas/cy}.
\end{equation}

\begin{figure}[hbt]
\begin{center}
\includegraphics[width=0.9\linewidth]{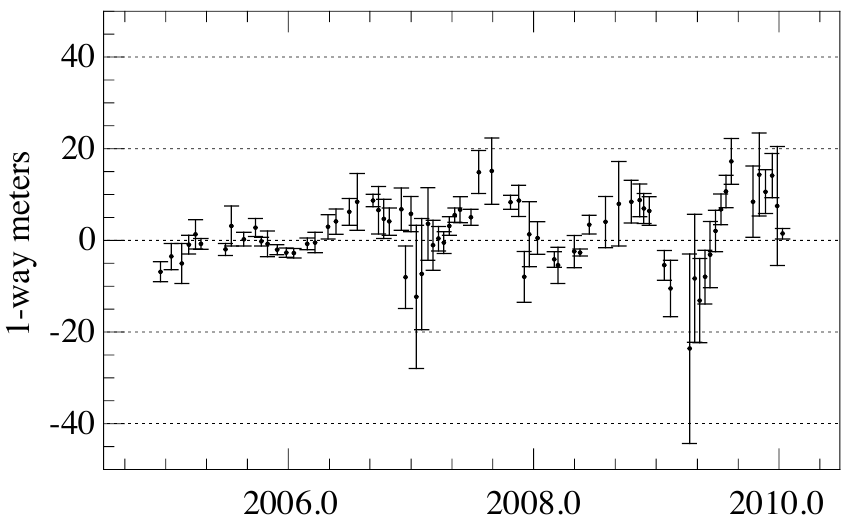}
\end{center}
\caption{Range residuals previously available.}
\label{fig:old}
\end{figure}

Other estimates on the anomalous perihelion precession of Saturn are given by the INPOP and EPM ephemerides \cite{fienga:2011qf,pitjeva:2010ys}. Nevertheless, these estimates were based on a preliminary reduction of the early Cassini range measurements considered here. The preliminary reduction was based on spacecraft orbits that were fit to both the Doppler and range measurements~\cite{folkner:2009fk,folkner:2010kx}. The corresponding range residuals are shown on Fig.~\ref{fig:old}. The remaining sinusoidal signature is an indication that the analysis was not satisfactory. Further analysis showed that the resulting spacecraft trajectories in separate orbits segments were not independent because of the use of the ranging data in the trajectory fits.  This led to the approach used here of fitting the spacecraft trajectories without use of the range data. 

\section{Conclusion}
In this paper, we used the Earth-Cassini range data in order to estimate the MOND external field effect parameter $Q_2$ that appears in MOND theory based on modified Poisson equation.
For this, we enhanced the dynamical model used for data analysis~\cite{standish:2012fk} to include the effects predicted by MOND given by~(\ref{eq:efe}).

The resulting estimate of $Q_2$ is given by~(\ref{eq:Q2est}). Cross validation tests show that the estimated uncertainty in this parameter is sound. No significant deviation from GR favoring MOND  has been detected in this set of data. Moreover, the very stringent constraint (\ref{eq:Q2est}) excludes the major part of the interval~(\ref{eq:Q2}) on $Q_2$ computed theoretically in \cite{blanchet:2011ys}. The constraint on $Q_2$ can be written as a constraint on $q$ (\ref{eq:qest1}-\ref{eq:qest2}) which can directly be compared to the Table I of \cite{milgrom:2009vn}. As a consequence, a large part of MOND theories based on modified Poisson equation and characterized by standard interpolating functions is excluded by Cassini range data (in particular the original TeVeS interpolating function is excluded by our analysis). 


\begin{acknowledgments}
The authors thank the anonymous referees for their useful comments, the Cassini mission operations team, the DSN staff, the radiometric data conditioning team and the media calibration team. The research described in this paper was carried out at the Jet Propulsion Laboratory, California Institute of Technology, under contract with the National Aeronautics and Space Administration \copyright\ 2013. A.H. thanks P. Wolf, B. Famaey and L. Iorio  for useful comments on this manuscript and acknowledges support from the Belgian American Educational Foundation (BAEF) and from the Gustave-Bo\"el - Sofina "Plateforme pour l'Education et le Talent".
\end{acknowledgments}

\bibliographystyle{apsrev4-1}

\bibliography{../../biblio}

\end{document}